\documentclass{aa}

\usepackage[varg]{txfonts}
\usepackage{natbib}
\usepackage{aalongtable}

\newcommand{\hm}{H$_2$} 

\begin{document}

\title{\hm\ formation on PAHs in photodissociation regions: a high-temperature pathway to molecular hydrogen}

\author{
  L. Boschman\inst{\ref{kapt},\ref{kvi}}\and
  S. Cazaux\inst{\ref{kapt}}\and
  M. Spaans\inst{\ref{kapt}}\and
  R. Hoekstra\inst{\ref{kvi}}\and
  T. Schlath\"{o}lter\inst{\ref{kvi}}
}

\institute{
Kapteyn Astronomical Institute, {University of Groningen}, P.O. Box 800, 9700 AV Groningen, the Netherlands, \email{boschman@astro.rug.nl} \label{kapt}
\and
Zernike Institute for Advanced Materials, {University of Groningen}, Nijenborgh 4, 9747 AG Groningen, the Netherlands \label{kvi}
}

\titlerunning{Hydrogenated PAHs in PDRs}
\authorrunning{Boschman et al.}

\abstract{}
{Molecular hydrogen is the most abundant molecule in the Universe.
It is thought that a large portion of \hm\ forms by association of hydrogen atoms to polycyclic aromatic hydrocarbons (PAHs).
We model the influence of PAHs on total \hm\ formation rates in photodissociation regions (PDRs) and assess the effect of these formation rates on the total cloud structure.}
{We set up a chemical kinetic model at steady state in a PDR environment and included radiative transfer to calculate the chemistry at different depths in the PDR.
This model includes known dust grain chemistry for the formation of \hm\ and a \hm\ formation mechanism on PAHs.
Since \hm\ formation on PAHs is impeded by thermal barriers, this pathway is only efficient at higher temperatures (T $>$ 200 K).
At these temperatures the conventional route of \hm\ formation via H atoms physisorbed on dust grains is no longer feasible, so the PAH mechanism enlarges the region where \hm\ formation is possible.}
{We find that PAHs have a significant influence on the structure of PDRs.
The extinction at which the transition from atomic to molecular hydrogen occurs strongly depends on the presence of PAHs, especially for PDRs with a strong external radiation field.
A sharp spatial transition between fully dehydrogenated PAHs on the outside of the cloud and normally hydrogenated PAHs on the inside is found.
As a proof of concept, we use coronene to show that \hm\ forms very efficiently on PAHs, and that this process can reproduce the  high \hm\ formation rates derived in several PDRs.
}{}


\maketitle

\section{Introduction}
\label{sec:introduction}

Molecular hydrogen is the most abundant molecule in the Universe \citep{shull1982, black1987, draine1996} and the main constituent of regions where stars are forming.
Many ion-molecule and neutral-neutral reactions are driven by the presence of \hm .
Specifically, \hm\ facilitates the formation of molecules such
as CO, OH, H$_{2}$O, HCO$^+$, HCN, and H$_3^+$ \citep{oppenheimer1974, dishoeck1986, tielens1985}.
These species, as \hm\ itself, can be important coolants and provide useful diagnostics of (irradiated) interstellar clouds and shocks \citep{neufeld1993, neufeld1995, lebourlot1999}.
As such, it is crucial to understand the formation of \hm\ from low ($\sim 10$ K) to high ($\sim 10^3$ K) temperatures.

Since gas phase routes to form \hm\ were found to be inefficient, dust grains are recognized as the favoured habitat for \hm\ formation \citep{oort1946, gould1963}.
These dust grains can adsorb hydrogen atoms on their surface with either a Van der Waals bond \citep[physisorption, ][]{pirronello1997a, pirronello1999, perry2003} or a covalent chemical bond \citep[chemisorption, ][]{zecho2002, hornekaer2006, mennella2006}.
The typical physisorption energy of a hydrogen atom is 700 - 800 K (60 - 70 meV), depending on the dust composition, whereas the typical chemisorption energy is 7000 - 25000 K (0.60 - 2.15 eV)  \citep{duley1993, cazaux2009}.
Because of their lower binding energies, physisorbed H atoms can easily migrate to another site (physisorbed or chemisorbed), or can evaporate back into the gas phase.
If a migrating atom encounters another adsorbed H atom, these two atoms can react to form \hm , the Langmuir-Hinshelwood mechanism (LH).
\hm\ can also be formed when a gas phase atom adsorbs onto an already occupied site (Eley-Rideal mechanism, ER).

For both the LH and the ER mechanism to work, hydrogen atoms must be able to adsorb and stick to the surface of the dust grain.
As the grain temperature rises, physisorbed H atoms start to evaporate and only chemisorbed H atoms are left on the grain surface.
This decreases the \hm\ formation rate, since the LH mechanism becomes inefficient and \hm\ formation relies on the ER mechanism involving chemisorbed H atoms.
The temperature at which the change of efficiency occurs depends on the physisorption energy, which depends on the composition of the dust grain.
Typically, physisorbed H atoms desorb at grain temperatures of 10 - 20 K \citep{hollenbach1971, cuppen2005, cazaux2009}.

However, high \hm\ formation rates have been observed in PDRs with high dust and gas temperatures, such that the observed rates cannot be explained by dust grains alone \citep{habart2004, allers2005}.
\citet{allers2005} modelled the structure of the Orion Bar PDR and matched it with observations of different \hm\ rovibrational lines.
Their best-fit model requires an \hm\ formation rate on dust grains of $3.8 \times 10^{-17} \mathrm{cm^{3} \ s^{-1}}$ at a temperature of 1000 K, which significantly exceeds the standard values for \hm\ formation on dust grains at 1000 K.
Furthermore, the dust grain FUV extinction cross section must be reduced by a factor of $\sim 3$ while the photoelectric heating rate in the atomic zone requires an increase of a factor of $\sim 3$ with respect to the a priori estimates.

\citet{habart2004} resolved these deviations from the canonical \hm\ formation rate coefficient \citep{jura1974} by including \hm\ formation on very small grains and polycyclic aromatic hydrocarbons (PAHs), and adjusting the barriers for \hm\ formation on dust grains.
Their model does not fully reproduce the observed \hm\ formation rates in warm, high-UV environments, such as $\rho$ Oph W and the Orion Bar, where rate coefficients as high as $10^{-16} \mathrm{cm^{3} \ s^{-1}}$ have been observed.

PAHs make up the low-mass end of the grain size distribution \citep{weingartner2001a}.
The bulk infrared emission features of PAHs have been detected in interstellar spectra \citep{allamandola1989}, but it is still impossible to identify individual molecules \citep{pilleri2009}.
Although PAHs are much smaller than the rest of the dust grains, they account for approximately half of the total dust grain surface area available for chemistry \citep{weingartner2001}.
In the past decades a plethora of studies has been performed to understand how these PAHs can catalyze the formation of molecular hydrogen.

One of the most widely used mechanisms to form \hm\ on PAHs consists of two different reactions, in which a gas phase hydrogen atom forms a covalent bond with a PAH and a second step where a different H atom abstracts a hydrogen atom from the PAH to form molecular hydrogen \citep{bauschlicher1998}.
Both reactions have been studied theoretically as well as experimentally with systems as large as coronene ($\mathrm{C_{24}H_{12}}$).
\citet{mennella2012} and \citet{thrower2012} exposed coronene films to atomic hydrogen and deuterium to derive cross sections for both the addition to and abstraction from coronene.
Density functional theory (DFT) calculations by \citet{rauls2008} indicated the presence of barriers when hydrogenating a PAH, which were later experimentally observed by \citet{boschman2012}.
These thermal barriers, which have a typical height of a few 100 K (a few times 10 meV), significantly restrict the formation of \hm\ on PAHs at low temperatures ($T \lesssim 200 \mathrm{K}$).

Alternatively, PAH molecules may lose an \hm\ molecule after being excited by a UV photon.
This photodesorption mechanism has been studied both theoretically and in the laboratory \citep{allain1996a, jochims1994, lepage2001}.
These studies find that loss of an H atom is the most dominant loss process.
However, \hm\ loss is only a few times smaller than H loss, which could make the former process an important candidate for \hm\ formation on PAHs, on the condition that H atoms also hydrogenate the PAH molecules.
It should be noted that the relative importance of these processes probably depends on the size of the PAH molecule.

Several theoretical models have been developed to quantify the formation of \hm\ on PAHs in astrophysical environments.
These models account for the formation of \hm\ on PAHs in different ways.
The Meudon PDR code by \citet{lepetit2006} considers PAHs as a small-sized extension of the dust grain size distribution, where PAH are treated as dust grains with a radius between 1 and 50 \AA .
A model by \citet{lepage2009} does treat the chemical reactions between H atoms and PAHs explicitly.
However, these reactions are limited to the normal hydrogenation state and a singly dehydrogenated state for both neutral and singly ionized molecules.
Additionally, it considers a singly superhydrogenated state for PAH cations.
\citet{montillaud2013} presented a model that studies the hydrogenation and charge states of PAHs in PDRs using the Meudon PDR code, where the possible hydrogenation states range from fully dehydrogenated up to the singly superhydrogenated state.
These hydrogenation states are available as either a neutral molecule or a singly ionized cation.
Although the PAH chemistry is different in \citet{montillaud2013} and \citet{lepage2009}, they both find that PAHs only exist in either the highest or the lowest possible hydrogenation state, depending on the column density.
Intermediate PAHs are rarely observed.

In this study we model the chemistry of PAHs in photodissociation regions (PDRs) to investigate their ability to form \hm .
We combine \hm\ formation through LH and ER on dust grains and through abstraction and photodesorption on PAHs to establish total \hm\ formation rates at different depths in a PDR.
We vary the number density of the cloud and the intensity of the external radiation field of the PDR.
This allows us to study the effect of the radiation field on the PAH chemistry.

We use a time-dependent rate equation model and report the results obtained at steady state.
Since addition barriers have only been derived for coronene, we use this molecule as our generic PAH.
The transferability of these results is discussed in Sect. \ref{ssec:imlar}.
We also incorporate the barriers determined in \citet{boschman2012} for the addition and abstraction of H atoms to coronene, which prohibits the chemical hydrogenation or abstraction cycle for \hm\ formation on PAHs at gas temperatures below 200 K.
In this manner, we find a high-temperature pathway for the formation of \hm .

In Sect. \ref{sec:model} we describe the details of the physics and chemistry of our model.
The results of the model are shown in Sect. \ref{sec:results}.
The uncertainties of the model are discussed in Sect. \ref{sec:discussion}, where we also compare our results and previous studies.

\section{Model}
\label{sec:model}
We describe a cloud by a one-dimensional PDR model with constant density $n_{\mathrm{H}}$.
The cloud is irradiated from one side by a star with a UV intensity field $G_{0}$, denoting the strength of the radiation field between $6 < h\nu < 13.6$ eV in Habing units \citep[$1 \ G_{0} = 1.6 \cdot 10^{-3} \ \mathrm{erg \ cm^{-2} \ s^{-1}}$,][]{habing1968}.
The cloud is divided into equally sized parcels of gas, which are consecutively brought into chemical equilibrium.
The photoprocesses that are included are photoionization of PAHs and carbon atoms, photodissociation of \hm\ , and photodesorption of H and \hm\ from PAHs.
All photoprocesses are mitigated by dust extinction at increasing column densities, and for \hm\ photodissociation we also include \hm\ self-shielding.
Ionization processes are counteracted by electron recombination, for which we use rates from \citet{ruiterkamp2005} and the online database KIDA \citep[\url{http://kida.obs.u-bordeaux1.fr}]{wakelam2012}.
The formation of \hm\ on interstellar dust grains (silicates and amorphous carbon) as well as on PAHs is considered in our model.

After a parcel has reached chemical equilibrium, we correct the radiation field for dust absorption and \hm\ self-shielding for the next parcel.
The process is then repeated until the entire cloud is in chemical equilibrium.
Dust grain and coronene abundances scale linearly with the number density.
The parameters defining the cloud are summarized in Table \ref{tab:parameters}.

\begin{table}
  \caption{Model parameters\label{tab:parameters}}
  \centering
  \begin{tabular}{l c c}
  \hline \hline
  Parameter & Definition & Value(s) \\
  \hline
  $G_{0}$ & radiation field & $10^{1.0} - 10^{5.5}$\\
  $n_{\mathrm{H}}$ & number density& $10^{2} - 10^{7} \mathrm{cm^{-3}}$\\
  $\sigma_{\mathrm{add}}$ & hydrogenation cross section & 1.10 $\mbox{\AA}^{2}$ / C atom \\
  $\sigma_{\mathrm{abs}}$ & abstraction cross section & 0.06 $\mbox{\AA}^{2}$ / C atom \\
  $E_{add}$ & hydrogenation barrier & see Table \ref{tab:reactions} \\
  $E_{abs}$ & abstraction barrier & 10 meV \\
  $N_{H,max}$ & maximum column density & $10^{23} \mathrm{cm^{-2}}$ \\
  $k_{\mathrm{ph}}$ & photodesorption rate & see Table \ref{tab:rates} \\
  \hline
  \end{tabular}
\end{table}

The processes involving PAHs that we consider are detailed in the following subsections.
In Sect. \ref{sssec:hydabs} we describe the processes of hydrogenation and abstraction of H atoms on PAHs.

In Sect. \ref{sssec:photdes} we discuss the interactions of PAHs with the impinging radiation field, where we describe the process of photoionization and electron recombination.
We also describe the process of photodesorption, where a hydrogen atom or an \hm\ molecule is removed from the PAH upon absorption of a photon.

The influence of dust grains on the PDR is described in Sect. \ref{ssec:dustgrains}, and we treat the photodissociation of \hm\ in Sect. \ref{ssec:molh}.
In Sect. \ref{ssec:comp} we discuss the model calculations themselves.

\subsection{PAHs}
\label{sssec:pahchem}
The abundance of PAHs in the interstellar medium (ISM) is very uncertain, and we refer to \citet{weingartner2001a} for a comparison between different values in several environments.
We assume a total carbon abundance of $2.5 \cdot 10^{-4}$, of which a third is locked in PAHs.
This is similar to studies by \citet{joblin1992} and \citet{li2001}, who find the fraction of carbon atoms locked in PAHs relative to hydrogen to be $6 \times 10^{-5}$.

We define PAHs as being smaller than 100 \AA, and we use coronene ($\mathrm{C_{24}H_{12}}$) as our prototypical PAH of the ISM.
This is motivated by the fact that coronene is a well-studied molecule, both theoretically and experimentally.
The barriers for single H addition and abstraction, which are instrumental for this study, have been determined only recently.
In this sense, we use coronene as a molecule representative of all the PAHs in the PDRs studied in this work.
It should be noted that this is a rough assumption, since it is widely believed that small PAHs ($\mathrm{N_{C}} \lesssim 40$) cannot survive in strong radiation fields, which would exclude the existence of the coronene molecule in PDR environments \citep{allain1996a}, and that compact PAHs are more stable than extended molecules \citep{jochims1994, ekern1997}.
However, other studies show that small PAHs can survive strong radiation fields, but in a very dehydrogenated state \citep{montillaud2013}.
Therefore a more realistic PDR model should consider a distribution of PAHs with different sizes, where larger PAHs with fewer H-bearing outer edge carbon atoms would be present.
This implies that these PAHs would be slightly less efficient in forming \hm , with an efficiency depending on the number of inner carbon atoms compared to outer edge carbon atoms. 
Although this would not affect the results drawn in this work, in future work a size distribution of PAHs would further increase the reliability of our model.
In Sect. \ref{ssec:imlar} we discuss the expected results when coronene is replaced with such a size distribution.

\subsubsection{Hydrogenation and abstraction}
\label{sssec:hydabs}
Hydrogen addition and subsequent abstraction constitute a two-step catalytic cycle for the formation of \hm .
First, a coronene molecule reacts with a hydrogen atom to form a singly superhydrogenated coronene molecule ($\mathrm{C_{24}H_{13}}$).
In the second reaction, another hydrogen atom reacts with this singly superhydrogenated coronene molecule, yielding molecular hydrogen and the coronene molecule \citep{bauschlicher1998}.

This process is not limited to fully aromatic coronene ($\mathrm{C_{24}H_{12}}$), but can also occur for coronene in other hydrogenation states, with up to 24 additional hydrogen atoms attached to the coronene ($\mathrm{C_{24}H_{36}}$).

A (hydrogenated) coronene molecule can be either neutral or positively or negatively charged, which significantly changes the chemistry.
We do not consider the PAH anions since their contributions are mostly important in strongly shielded environments, but concentrate on the barely shielded edges of the PDRs.
The system of chemical reactions between coronene and hydrogen is summarized in Eqs. \ref{eq:genhyd} and \ref{eq:genabs} and is detailed in Table \ref{tab:reactions}.

\begin{eqnarray}
  \label{eq:genhyd}
  \mathrm{C_{24}H_{n}^{0/+}} + \mathrm{H} &\rightarrow& \mathrm{C_{24}H_{n+1}^{0/+}}  \\
  \label{eq:genabs}
  \mathrm{C_{24}H_{n+1}^{0/+}} + \mathrm{H} &\rightarrow& \mathrm{C_{24}H_{n}^{0/+}} + \mathrm{H_{2}}.
\end{eqnarray}

These reactions take place at a rate $r$, which is calculated as 
\begin{equation}
  \label{eq:rate}
  r(\mathrm{n})  = n \left(\mathrm{H}\right) n \left( \mathrm{C_{24}H_{n}^{0/+}}\right) \sigma v_{\mathrm{th}} \exp \left( - \frac{E_{a}}{k_{B} T} \right).
\end{equation}

Here, $n \left(\mathrm{H}\right)$ and $n \left(\mathrm{C_{24}H_{n}^{0/+}} \right)$ represent the number densities of atomic hydrogen and coronene molecules or cations containing n H atoms, respectively.
$\sigma$ represents the cross section for the specific reaction and $v_{\mathrm{th}}$ is the average thermal velocity of atomic hydrogen.
We use a cross section of 1.1 $\mathrm{\AA^{2}}$ per reactive carbon atom for hydrogenation and 0.06 $\mathrm{\AA^{2}}$ per reactive carbon atom for abstraction \citep{mennella2012}.

Since coronene is much more massive than hydrogen, its velocity is neglegible compared to that of the gas phase hydrogen atoms.
$v_{\mathrm{th}}$ is therefore the thermal velocity of hydrogen atoms, which is calculated as

\begin{equation}
  \label{eq:vtherm}
  v_{\mathrm{th}} = \sqrt{\frac{8 k_{B} T}{\pi m_{\mathrm{H}}}},
\end{equation}
where $\mathrm{k_{B}}$ is the Boltzmann constant, $T$ is the gas temperature and $m_{\mathrm{H}}$ is the mass of the hydrogen atom.

The hydrogenation and abstraction of coronene is subject to barriers ($E_{a}$), depending on hydrogenation and charge state.
The presence of these barriers was predicted for neutral coronene by \citet{rauls2008} and first experimentally determined for coronene cations by \citet{boschman2012}.
The latter study shows an alternating behaviour for the existence of barriers, meaning that the first hydrogenation of the coronene cation is barrierless.
The second hydrogenation then has a barrier, while the third does not, etc.
More recent experiments in conjuntion with theoretical calculations \citep{boschman2014} show that this alternating behaviour continues until the coronene cation is fully hydrogenated.
The hydrogenation barriers are summarized in Table \ref{tab:reactions}.

Abstraction barriers are only present for abstraction from those hydrogenation states for which adding a hydrogen atom would also have a barrier.
In this manner, a distinction is made between chemically stable and chemically unstable hydrogenation states, where the unstable ones have a radical atom.
The barrier against abstraction is set at 10 meV, equal to the value calculated by \citet{rauls2008} for abstracting an outer edge H atom from the doubly hydrogenated coronene molecule ($\mathrm{C_{24}H_{14}}$).
The influence of the uncertainties of these barrier heights on the obtained results is discussed in Sect. \ref{sec:discussion}.

With the mechanism for \hm\ formation catalyzed by PAHs as described in Eqs. \eqref{eq:genhyd} and \eqref{eq:genabs}, the total \hm\ formation rate due to abstraction from PAHs is calculated as
\begin{equation}
  \label{eq:pahh2form}
  R_{f} (\mathrm{PAH_{abs}}) = \sum_{\mathrm{n}=0}^{36} r(\mathrm{n}),
\end{equation}
where $r(\mathrm{n})$ is the \hm\ formation rate as shown in Eq. \eqref{eq:rate}.

\subsubsection{Photodesorption}
\label{sssec:photdes}
In a PDR, PAHs experience a strong interaction with the photons from the impinging radiation field, which can either ionize the PAH or excite it into a higher state.
We use Eq. \eqref{eq:ionrate} from \citet{lepage2001} to calculate the total photoionization rate:
\begin{equation}
  \label{eq:ionrate}
  k_{\mathrm{ion}} = \int_{\mathrm{IP}}^{13.6 \ \mathrm{eV}} Y_{\mathrm{ion}}(E) \sigma_{\mathrm{UV}} (E) N(E) d E,
\end{equation}
with IP the ionization potential of the molecule and $Y_{\mathrm{ion}}$ the ionization yield taken from \citet{lepage2001}. $\sigma_{\mathrm{UV}} (E)$ is the photoabsorption cross section \citep{malloci2007} and $N(E)$ is the intensity of the Habing interstellar radiation field in photons $\mathrm{cm^{-2} s^{-1} eV^{-1}}$ \citep{draine1978}.

If the molecule is not ionized but excited, it can dissipate the excitation energy by either IR emission, loss of an H atom, or loss of an \hm\ molecule \citep{allain1996a}.
To calculate the rates for these processes, we use the RRKM method as described in \citet{lepage2001}, which gives the rate of bond dissociation as a function of excitation energy, bond energy and vibrational degrees of freedom in the molecule.
The competition between processes is accounted for by dividing the rate of one process by the sum of the rates of all possible processes.
In this way, the branching ratio for each dissociation pathway upon excitation is obtained.
Multiplying this dissociation probability with the coronene photoabsorption cross section $\sigma_{\mathrm{UV}} (E)$ of \citet{malloci2007} and integrating over the Habing interstellar radiation field $N(E)$ below the Lyman limit gives the total photodissociation rate per $G_{0}$, as shown in Eqs. \eqref{eq:phdes_h} and \eqref{eq:phdes_h2}:
\begin{align}
  \label{eq:phdes_h}
  k_{\mathrm{ph}}(\mathrm{H}) &= \int_{0}^{13.6 \mathrm{\ eV}} \left( 1 - Y_{\mathrm{ion}}  \right) \frac{k_{\mathrm{diss}}(\mathrm{H})}{k_{\mathrm{diss}}(\mathrm{H}) + k_{\mathrm{rad}}' + k_{\mathrm{diss}}(\mathrm{H_{2}})} \nonumber \\
  & \qquad \qquad \qquad {} \times \sigma_{\mathrm{UV}} (E) N(E) d E ;\\
  \label{eq:phdes_h2}
  k_{\mathrm{ph}}(\mathrm{H_{2}}) &= \int_{0}^{13.6 \mathrm{\ eV}} \left( 1 - Y_{\mathrm{ion}}  \right) \frac{k_{\mathrm{diss}}(\mathrm{H_{2}})}{k_{\mathrm{diss}}(\mathrm{H}) + k_{\mathrm{rad}}' + k_{\mathrm{diss}}(\mathrm{H_{2}})} \nonumber \\
  & \qquad \qquad \qquad {} \times \sigma_{\mathrm{UV}} (E) N(E) d E,
\end{align}
where $k_{\mathrm{rad}}'$ is the rate of IR stabilization.

We use binding energies from \citet{lepage2001} and \citet{paris2014} to calculate the photodissociation rates for both the neutral coronene and the coronene cation in their dehydrogenated, superhydrogenated, and normally hydrogenated states.
These photodissociation rates cover both loss of a single H atom and the loss of \hm\ and are summarized in Table \ref{tab:rates}.

In addition to PAH ionization, we also include the photoionization of unbound carbon atoms as a source of free electrons.
Recombination reactions are included in our chemical network, which ensures an ionization balance.
The rates for recombination reactions on PAH cations are given by
\begin{equation}
  k_{\mathrm{rec}} = 3 \times 10^{-7}  \left( \frac{300}{T} \right)^{1/2} s \left( \mathrm{e} \right) n_{\mathrm{e}},
\end{equation}
with $T$ the temperature, $s \left( \mathrm{e} \right)$ the electron sticking coefficient and $n_{\mathrm{e}}$ the number density of free electrons \citep{ruiterkamp2005}.
From \citet{ruiterkamp2005} we adopt an electron sticking coefficient of $s \left( \mathrm{e} \right) = 10^{-3}$ for coronene.

We only use carbon atoms as a source of free electrons, because other metals with an ionization energy comparable to carbon have abundances that are at least one order of magnitude lower.
Since the ionization balance is driven by radiation, other metals are not considered.

\subsection{\hm\ formation on dust grains}
\label{ssec:dustgrains}
Dust grains have a significant influence on the chemistry taking place in the PDR because of their catalyzing properties, which strongly depend on the chemical composition of the dust grain, for example, silicates or amorphous carbon.

In our model, we limit the dust grain chemistry to the catalysis of \hm\ formation, and for this we assume the first H atom to be either physisorbed or chemisorbed.
There are then two mechanisms for this H atom to react with another H atom and to form \hm .
In the first process, an H atom from the gas phase interacts with an adsorbed H atom (Eley-Rideal).
In the second process, a different physisorbed H atom travelling around the surface interacts with the adsorbed H atom (Langmuir-Hinshelwood).

For both mechanisms to occur, a hydrogen atom must first adsorb onto the surface of the dust grain and become either chemisorbed or physisorbed.
This adsorption process is incorporated into rate equations with a temperature-dependent sticking coefficient \citep{burke1983}.
Once an H atom is adsorbed onto the surface, it has a certain probability to form \hm .
This probability $\epsilon$ strongly depends on the binding energies of the adsorbed atoms and the gas and dust temperatures.
The resulting \hm\ formation rate equation is given by
\begin{equation}
  \label{eq:dustform}
  r_{\mathrm{grain}} = n \left( \mathrm{grain} \right) n \left( \mathrm{H} \right) v_{\mathrm{th}} \sigma \epsilon S .
\end{equation}
Here $n \left( \mathrm{grain} \right)$ is the number density of dust grains, $n \left( \mathrm{H} \right)$ is the number density of atomic hydrogen, and $v_{\mathrm{th}}$ is the average thermal velocity in the gas phase.
$\sigma$ is the average cross section of a dust grain, and values for $n \left( \mathrm{grain} \right) \sigma$ are taken from \citet{cazaux2009}.
For the values of formation efficiency $\epsilon$ and sticking coefficient $S$, we use the expressions of \citet{cazaux2009} and the dust grain size distribution derived by \citet{weingartner2001a}.
The sticking coefficient is calculated as
\begin{align}
  S(T_{gas} , T_{dust}) &= \left(1 + 0.4 \cdot \left( \frac{T_{gas} + T_{dust}}{100} \right)^{0.5} \right.\nonumber\\
  & \qquad \qquad \left. + \ 0.2 \cdot \frac{T_{gas}}{100} + 0.08 \cdot \left( \frac{T_{gas}}{100} \right)^{2} \right)^{-1},
\end{align}
which gives a numerical value for $S$ typically between 0.4 and 0.9.
The formation efficiency strongly depends on the surface composition and has a typical value between 0.4 and 1 for amorphous carbon;
it lies between 0.01 and 1 for silicates \citep{cazaux2009}.

\citet{cazaux2009} considered \hm\ formation on the surfaces of silicates, amorphous carbon and PAHs.
We avoid a redundancy by only taking into account \hm\ formation on silicates and amorphous carbon and defining this as \hm\ formation on dust.
The \hm\ formation on dust is calculated by using the surface area of silicates ($\frac{n_{gr} \sigma}{n_{H}} = 10^{-21} \mathrm{cm^{-2}}$) and amorphous carbon ($\frac{n_{gr} \sigma}{n_{H}} = 1.7 \times 10^{-22} \mathrm{cm^{-2}}$), so as to not count the contribution of PAHs twice.

\subsection{H$_{2}$ photodissociation}
\label{ssec:molh}

The destruction mechanism for molecular hydrogen in the presence of UV photons is dissociation subsequent to the photo-excitation into a higher electronic state.
Approximately 10\% of these excited \hm\ molecules undergo fluorescence into the vibrational continuum of the electronic ground state, giving rise to dissociation of the molecule.
The \hm\ photodissociation rate is given by \citet{black1987} and \citet{draine1996}:

\begin{equation}
  \label{eq:h2diss}
  k_{\mathrm{diss}} = 3.4 \cdot 10^{-11} f( N( \mathrm{H_{2}} ) ) G_{0} \exp \left[ -2.6 A_{V} \right] \mathrm{s^{-1}}.
\end{equation}
In this equation the exponential factor expresses the dust extinction.
The factor $f( N( \mathrm{H_{2}} ) )$ accounts for \hm\ self-shielding against photodissociation.
This self-shielding factor $f( N( \mathrm{H_{2}} ) )$ has been computed as a function of \hm\ column density by \citet{draine1996},

\begin{eqnarray}
  \label{eq:h2shield}
  f( N( \mathrm{H_{2}} ) ) = \frac{0.965}{\left( 1 + x/b_{5} \right)^{2}} &+& \frac{0.035}{\sqrt{1 + x}} \\
  \nonumber
  &\times& \exp \left[ -8.5 \cdot 10^{-4} \sqrt{1 + x} \right],
\end{eqnarray}
where $x = N(\mathrm{H_{2}}) / 5 \cdot 10^{14} \ \mathrm{cm^{-2}}$ is the \hm\ column density and $b_{5}$ is the line width of \hm\ absorption lines expressed in $10^{5} \mathrm{cm \ s^{-1}}$, which was set to unity for our calculations.

\subsection{Computations}
\label{ssec:comp}

In our calculations, we use a one-dimensional plane-parallel slab of gas that is irradiated from one side.
This slab is divided into $10^{3}$ equally sized cells of gas, which all have the same number density and abundances.
Initially, there is no molecular hydrogen present, and the only PAH is the normally hydrogenated coronene molecule ($\mathrm{C_{24}H_{12}}$).

Every single reaction in the list of reactions gives rise to a rate equation, and from this the rate for every reaction is calculated at $t=0$.
We perform a numerical Euler integration until the system reaches convergence, which is defined as the moment when the fractional difference between two time steps is smaller than $10^{-3}$ for every chemical species.
For the species with abundances higher than $10^{-9}$, this condition is met in less than $10^3$ years.
For species with lower abundances it is increasingly difficult to converge, and timescales ranging between $10^5$ and $10^8$ years are required to reach equilibrium for the entire system.
Time-dependent environmental effects may play a role, but this will be a topic of future study.
After convergence is reached in a parcel of gas, the calculation starts for the next parcel of gas, including a modification of the radiation field to account for dust extinction and \hm\ self-shielding.

Temperatures in a PDR are strongly dependent on the external radiation field through various mechanisms.
In regions where dust grains are exposed to UV radiation, the gas is indirectly heated via photoionization of dust grains and PAHs.
At low densities ($n_{\mathrm{H}} \lesssim 3 \cdot 10^{4} \mathrm{cm^{-3}}$), the thermal coupling between dust grains and the gas phase does not occur, which leads to gas temperatures significantly different from the dust temperatures.
At higher densities ($n_{\mathrm{H}} \gtrsim 3 \cdot 10^{4} \mathrm{cm^{-3}}$) thermal coupling between gas and dust grains has to be taken into account.
We extract gas and dust grain temperatures from models of PDRs (\citet{meijerink2005} and refined by \citet{hocuk2011}, available at \url{http://www.mpe.mpg.de/~seyit/Tgas/}).
These models derive gas and dust grain temperatures from the different heating and cooling processes, leading to temperatures for a given combination of radiation field, number density, and column density.
We use the outcomes of these models in tabular form, from which we extract the temperatures for our simulations.
Temperature fluctuations of dust grains are not considered in this model, because \citet{bron2014} showed that their effect on the \hm\ formation rate is limited.\\

The external radiation field has a significant influence on the structure of the cloud through photoreactions and heating of the gas.
At high total column densities ($N_{\mathrm{H}} \approx 10^{21} \ \mathrm{cm^{-2}}$) and at solar metallicity, the intensity of the radiation field diminishes due to dust absorption and scattering.
The radiation field is therefore a function of column density, as shown in Eq. \ref{eq:dustabs}.
\begin{equation}
  \label{eq:dustabs}
  G_{0} (A_{\mathrm{V}}) = G_{0} (0) \exp \left[ - \beta \ A_{\mathrm{V}} \right] .
\end{equation}
In this expression, $\beta$ is the process-dependent extinction factor, which has a typical value of 2.5, and $A_{\mathrm{V}}$ is the visual extinction calculated by $A_{\mathrm{V}} =  (N(\mathrm{H}) + 2 N(\mathrm{H_{2}})) / 2.2 \cdot 10^{21} \mathrm{cm^{-2}}$ \citep{guver2009}.

\section{Results}
\label{sec:results}

From our model we obtain number densities for all coronene charge and hydrogenation states, as well as the \hm\ formation rate associated with each of these species.
We also obtain the \hm\ formation rates on dust grains and the number densities of atomic and molecular hydrogen.
Together with the gas and dust grain temperatures, this gives us a clear view of the chemistry and physics inside the PDR.
We present the spatial distribution of coronene hydrogenation states in Sect. \ref{ssec:spatpahdis}.
In Sect. \ref{ssec:h2form} we show the \hm\ formation rates throughout the PDR, and in Sect. \ref{ssec:impact} we show how the formation of \hm\ on coronene impacts the PDR structure.

To study the hydrogen-coronene chemistry, the spatial distribution of coronene hydrogenation states is tracked.
A region with low hydrogenation states ($\mathrm{C_{24}^{0,+}}$ and $\mathrm{C_{24}H^{0,+}}$ dominate) indicates that photodesorption occurs at a rate significantly higher than that of hydrogen addition.
Conversely, in a region with normal and high hydrogenation states ($\mathrm{C_{24}H_{12}^{0,+}}$ and higher dominate), hydrogenation dominates H-loss processes.
Since \hm\ formation is possible through both photodissociation of coronene and chemical abstraction on coronene, the spatial distribution of the hydrogenation states of coronene can tell us which of these processes is more important for \hm\ formation.
Furthermore, in Sect. \ref{ssec:h2form} we compare the formation of \hm\ on coronene with the formation on dust grains, and report our results.

\subsection{Spatial distribution of coronene}
\label{ssec:spatpahdis}

\begin{figure}
  \resizebox{\hsize}{!}{
  \includegraphics{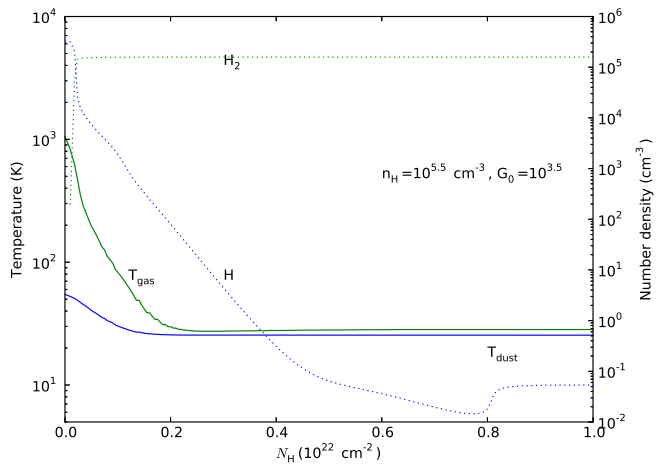}}
  \caption{Distribution of dust and gas temperatures throughout a gas cloud with $G_{0} = 10^{3.5}$ and $n_{\mathrm{H}} = 10^{5.5} \mathrm{cm^{-3}}$. The dotted lines show the distribution of  H and \hm .}
  \label{fig:pahtemp}
\end{figure}

The spatial distribution of coronene states is due to the temperature and radiation profiles of the cloud.
At the boundaries of the cloud, the radiation field is very strong and the gas temperature is high ($T_{\mathrm{gas}} > 200 K$), as is shown in Fig. \ref{fig:pahtemp}.
At these high gas temperatures coronene hydrogenation is very efficient because the H atoms can then easily overcome the barriers against hydrogenation.
However, this is easily offset by the intense radiation field that completely dehydrogenates coronene down to the carbon skeleton.
Deeper inside the cloud the radiation field is decreased, causing a decrease in the photoionization rate and a subsequent gradual transition from fully dehydrogenated coronene cations to their neutral counterparts is expected.
At even higher column densities, the radiation field is too weak for photodissociation to dominate.
As a result, a transition to the neutral, normally hydrogenated coronene molecule should occur.

\begin{figure}
  \resizebox{\hsize}{!}{
  \includegraphics{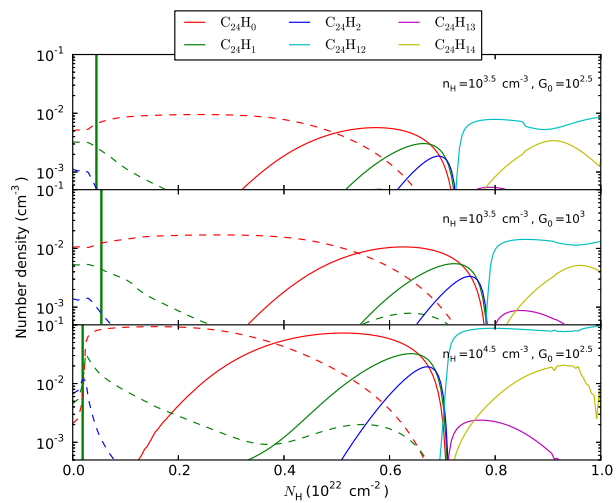}}
  \caption{Distribution of coronene charge and hydrogenation state for three different values of $G_{0}$ and $n_{\mathrm{H}}$. Coronene cations are plotted with dashed lines, and neutral coronene molecules have solid lines. Only the most abundant hydrogenation states are shown. The top graph has $G_{0} = 10^{2.5}$ and $n_{\mathrm{H}} = 10^{3.5} \mathrm{cm^{-3}}$, the middle graph $G_{0} = 10^{3.0}$ and $n_{\mathrm{H}} = 10^{3.5} \mathrm{cm^{-3}}$ , and the bottom graph corresponds to $G_{0} = 10^{2.5}$ and $n_{\mathrm{H}} = 10^{4.5} \mathrm{cm^{-3}}$. The vertical line marks the location of the H/\hm\ transition, where $n(\mathrm{H}) = 2 \cdot n(\mathrm{H_{2}})$.}
  \label{fig:pahdens}
\end{figure}

A typical spatial distribution of hydrogenated coronene is presented in Fig. \ref{fig:pahdens}.
In the outskirts of the cloud coronene is fully dehydrogenated and exists as cations.
Deeper inside the cloud, the dominant form of coronene is the fully dehydrogenated neutral molecule, which is subsequently overturned by the normally hydrogenated neutral coronene molecule.

Clearly absent in our results are the highly superhydrogenated species, although coronene with a few extra hydrogen atoms can be abundant at high extinctions.
This can be explained by the correlation between the UV intensity and the gas temperature.
A high gas temperature is necessary to overcome the energy barriers in the hydrogenation process.
These gas temperatures are only reached in regions where the impinging radiation field is strong enough to heat the gas to such temperatures.
At the same time, this radiation field is strong enough to photodissociate all the coronene molecules.
This is consistent with previous PDR models, such as those of \citet{montillaud2013}, who modelled the north-west PDR of NGC 7023.
They reported that at every depth there is one dominant hydrogenation state, which is either the one with the most H atoms removed or the highest possible hydrogenation state.
Their hydrogenation states range from full dehydrogenation to one superhydrogenation.
\citet{montillaud2013} concluded that coronene is fully dehydrogenated throughout the entire cloud ($A_{\mathrm{V}} \leq 3$ mag).
This is consistent with our findings, where we report that coronene is dehydrogenated at the edge of the cloud, but normally hydrogenated at extinctions of more than a few mag.

The results shown here are also consistent with a recent study that interpreted Spitzer and ISOCAM observations of NGC 7023 \citep{joblin2010}.
Using the method developed by \citet{pilleri2012} to identify PAH cations and neutral PAHs, \citet{joblin2010} showed that there is a sharp transition between PAH cations at the edge of the PDR and neutral PAH molecules deeper into the cloud.

The radiation field has a very clear impact on the coronene distribution through two different mechanisms, as shown in Fig. \ref{fig:pahdens}.
First, an increased radiation field leads to a higher \hm\ photodissociation rate, pushing the H/\hm\ front to higher column densities.
Second, a higher $G_{0}$ leads to a higher rate of photoionization of PAHs, changing the transition from cations to neutrals to higher column densities.
Furthermore, a stronger radiation field drives the transition from dehydrogenated to normally hydrogenated PAHs to higher column densities.

An increase in the number density of the cloud has the opposite effect of increasing the radiation field.
Hence, the hydrogenated/dehydrogenated front is shifted towards lower column densities, as shown in Fig. \ref{fig:pahdens}.

\subsection{\hm\ formation rates}
\label{ssec:h2form}

The \hm\ formation rates on coronene and dust grains are strongly dependent on the radiation field and the dust temperature, respectively.
For \hm\ formation on dust, the dust must be cold enough to prevent physisorbed H atoms from desorbing into the gas phase.
When $T_{\mathrm{dust}}$ rises above 15-20 K, \hm\ formation is only possible with chemisorbed H atoms and is less efficient.

For coronene, however, there are two possible pathways for \hm\ formation, as described in Sects. \ref{sssec:hydabs} and \ref{sssec:photdes}.
The dominant \hm\ formation process is the photodesorption of \hm\ from coronene with at least two hydrogen atoms.
Although the occurence of these molecules is low in PDRs, this process occurs very fast because of the intense radiation.
This results in a fast catalytic cycle where the bare carbon skeleton of the coronene molecule consecutively adsorbs two H atoms, which are subsequently photodesorbed as an \hm\ molecule.
The \hm\ formation rates for this mechanism can be as high as a few times $10^{-16} \mathrm{cm^{3} s^{-1}}$, which is similar to \hm\ formation rates in PDRs found by \citet{habart2004}.

The formation of \hm\ through chemical abstraction on coronene is not a dominant process in PDR environments, although the temperatures are high enough to easily cross the thermal barriers associated  with this process.
For the latter process hydrogen atoms must be adsorbed to the coronene molecule, but the high-radiation field photodesorbs almost all H atoms from the coronene molecule.
This leaves most coronene molecules without H atoms, and a small fraction has only one or two hydrogen atoms, rendering the chemical abstraction process negligible.

\begin{figure}
  \resizebox{\hsize}{!}{
  \includegraphics{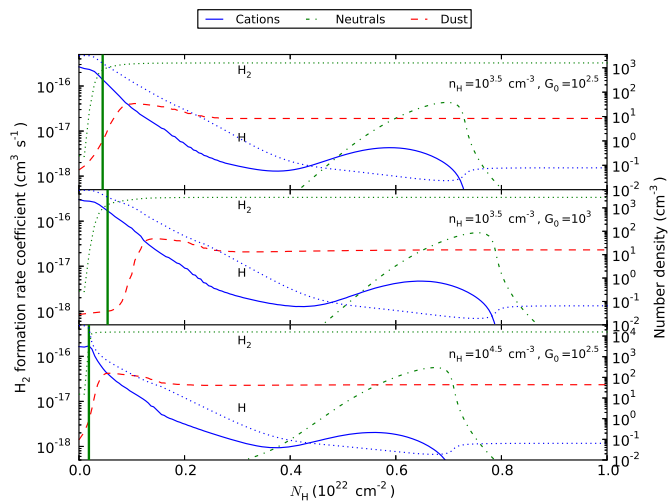}}
  \caption{\hm\ formation rate coefficients on cationic and neutral coronene and dust (solid, dashed-dotted, and dashed lines, left y-axis) and  distributions of H and \hm\ (dotted lines, right y-axis). The top graph has $G_{0} = 10^{2.5}$ and $n_{\mathrm{H}} = 10^{3.5} \mathrm{cm^{-3}}$, the middle graph $G_{0} = 10^{3.0}$ and $n_{\mathrm{H}} = 10^{3.5} \mathrm{cm^{-3}}$ , and the bottom graph corresponds to $G_{0} = 10^{2.5}$ and $n_{\mathrm{H}} = 10^{4.5} \mathrm{cm^{-3}}$. The vertical line marks the location of the H/\hm\ transition, where $n(\mathrm{H}) = 2 \cdot n(\mathrm{H_{2}})$.}
  \label{fig:form}
\end{figure}

To compare \hm\ formation on coronene with \hm\ formation on dust grains, we present the \hm\ formation rate coefficients as a function of column density in Figs. \ref{fig:form} and \ref{fig:pahstruc}.
In the same figures the number densities of atomic and molecular hydrogen are reported.
To study the effect of the radiation field and the total hydrogen number density, the simulations presented in the three panels of Fig. \ref{fig:form} differ in radiation field or number density $n_{\mathrm{H}}$.

At low column densities, the gas and dust are heated by the impinging radiation field, making coronene the dominant source of \hm\ formation.
At increasing column densities, the temperatures start to decline, and dust grains take over as the dominant \hm\ formation agent.

Increasing the radiation field heats the gas and the dust grains, making the transition from PAHs to dust grains occur at higher column densities, which is shown in the middle panel in Fig \ref{fig:form}.
Increasing the density has the opposite effect of increasing the radiation field, causing a decrease of the column density for the PAH/dust grain transition, as illustrated in the bottom panel of Fig. \ref{fig:form}.

\subsection{Impact of \hm\ formation on coronene on the PDR structure}
\label{ssec:impact}

\hm\ formation on coronene can significantly change the structure of a cloud.
This influence manifests itself in the column density of the H/\hm\ transition in the cloud.
We consider the H/\hm\ transition to occur at a column density where $2 n(\mathrm{H_{2}}) = n(\mathrm{H})$.
This column density is lowered by the presence of PAHs in the PDR.

\begin{figure}
  \resizebox{\hsize}{!}{
  \includegraphics{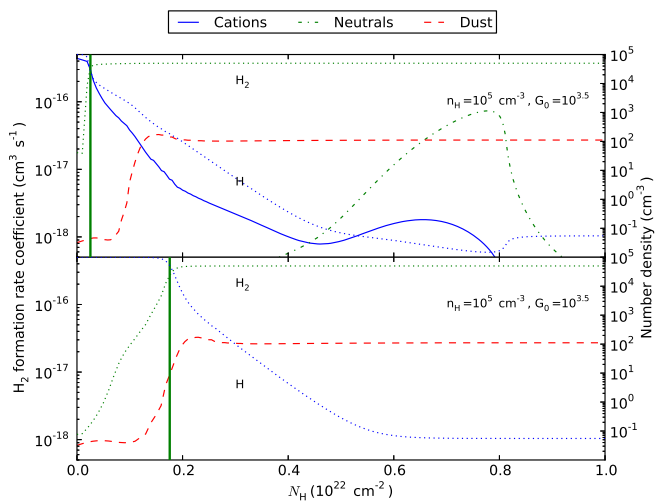}}
  \caption{Comparison between two identical clouds, with (top panel) and without (bottom panel) \hm\ formation on coronene. \hm\ formation rate coefficients on ionized (solid) and neutral (dashed-dotted) coronene and dust grains (dashed lines) are represented on the left y-axis. Number densities of H and \hm\ (dotted lines) are represented on the right y-axis. The thick vertical green line indicates the location where $n(\mathrm{H}) = 2 \cdot n(\mathrm{H_{2}})$. $G_{0} = 10^{3.5}$ and $n_{\mathrm{H}} = 10^{5} \mathrm{cm^{-3}}$.}
  \label{fig:pahstruc}
\end{figure}

\begin{figure}
  \resizebox{\hsize}{!}{
  \includegraphics{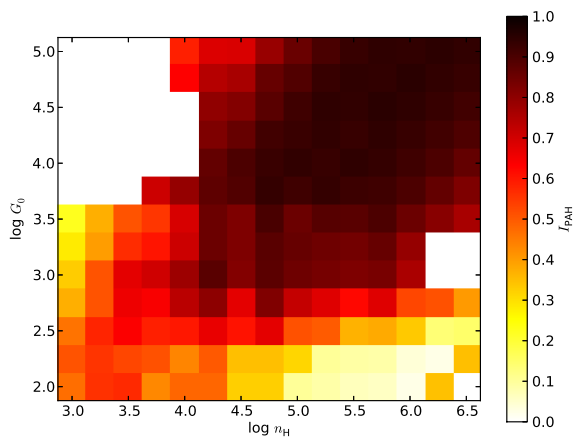}}
  \caption{PAH influence parameter plotted as a function of number density and external radiation field. The influence of PAHs on the cloud structure increases with increasing number density and radiation field. White regions indicate that no data are available.}
  \label{fig:influence}
\end{figure}

An example is presented in Fig. \ref{fig:pahstruc}, where we show the number density profiles of atomic and molecular hydrogen.
In the top panel of Fig. \ref{fig:pahstruc}, \hm\ can be formed by reactions on coronene and dust grains, and in the bottom panel only \hm\ formation on dust grains is considered.
Comparing these figures, we see that the column density of the H/\hm\ transition is lowered by a factor of 4 due to the formation of \hm\ on coronene.
The influence of coronene can be quantified by the ratio of the column densities of the transition with and without considering the formation of \hm\ on coronene.
We define the influence parameter $I_{\mathrm{PAH}}$ as
\begin{equation}
  I_{\mathrm{PAH}} = 1 - \frac{N \left( H \right)_{\mathrm{trans}}^{\mathrm{dust\ + \ PAHs}}}{N \left( H \right)_{\mathrm{trans}}^{\mathrm{dust}}} ,
\end{equation}
which ranges between 0 and 1.
In the case of a value of 0, coronene has no influence on the column density of the H/\hm\ transition.
When $I_{\mathrm{PAH}} = 1$, coronene shifts the H/\hm\ transition completely to the edge of the cloud.
The highest value of $I_{\mathrm{PAH}}$ present in our results is approximately 0.8.
At this value, the rate of \hm\ formation on coronene is $~ 200$ times higher than \hm\ formation on dust grains at the column density of the H/\hm\ transition, while these rates are approximately equal for the lowest values of the influence parameter.
This influence parameter is shown as a function of cloud number density and external radiation field in Fig. \ref{fig:influence}.
The influence parameter increases with both increasing number density and external radiation field.

An increased number density increases the rate of reactions between coronene and H atoms, resulting in a higher abundance of PAHs with one or more hydrogen atoms.
At a similar radiation field, this yields a higher \hm\ formation rate, pushing the H/\hm\ transition to lower column densities.
A similar argument applies to the dependence on the radiation field.
With an increasing radiation field, all photodissociation rates increase, including those of the \hm\ forming processes.
However, the effect is more subtle because a stronger radiation field changes the population distribution of the hydrogenation states towards a lower fraction of coronene with two or more H atoms.
This makes fewer coronene molecules eligible for the release of \hm\ upon photoabsorption.
This effect leads to an optimal radiation intensity for every number density for maximal coronene influence.
If the radiation field has a higher intensity, the influence of coronene on the position of the H/\hm\ front decreases again, as is shown in Fig. \ref{fig:influence}.

\begin{figure}
  \resizebox{\hsize}{!}{
  \includegraphics{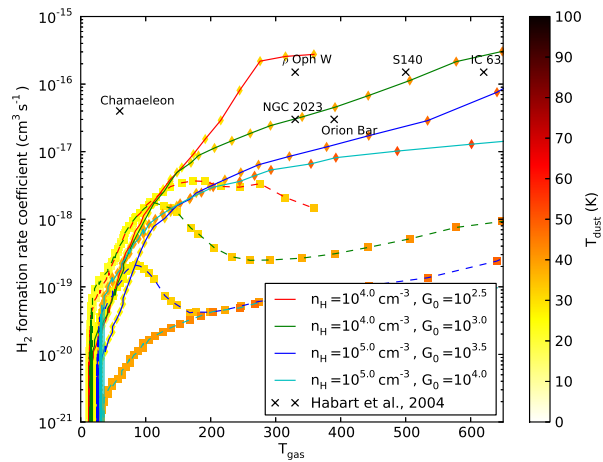}}
  \caption{\hm\ formation rates on for different values of $G_{0}$ and $n_{\mathrm{H}}$ as a function of $T_{\mathrm{gas}}$. Solid lines indicate \hm\ formation by PAHs and dashed lines represent dust grain \hm\ formation. For each datapoint, the color of the marker corresponds to the dust temperature of that datapoint. Both the gas and dust temperatures have been derived from the models in \citet{hocuk2011}. The black markers indicate the \hm\ formation rates reported by \citet{habart2004} for different PDRs. These formation rates have been estimated using \hm\ line intensity ratios from the \hm\ fluorescent lines.}
  \label{fig:h2form}
\end{figure}

Strong radiation fields also correlate with high dust grain temperatures, which inhibit \hm\ formation.
Figure \ref{fig:h2form} shows the \hm\ formation rate of dust grains (solid lines) and coronene (dashed lines) as a function of gas temperature for different values of $G_{0}$ and $n_{\mathrm{H}}$.
For each datapoint, the color of the marker indicates the dust temperature.
Additionally, observational \hm\ formation rates derived by \citet{habart2004} for different PDRs are reported in this figure.
It is clear that \hm\ formation on PAHs is necessary to reproduce the observed \hm\ formation rates.

From this plot it is also clear that dust grains and PAHs contribute to \hm\ formation in different temperature regimes.
Since the temperature decreases with increasing column density, PAHs have their strongest \hm\ formation activity at the edges of PDRs, while dust grains dominate this process deeper inside the cloud.
PAHs change the cloud structure the most when the conversion from H to \hm\ occurs at lower extinction than the one at which dust grains start to dominate \hm\ formation.
We can see from Fig. \ref{fig:h2form} that this is the case when the transition occurs in a region where $T_{\mathrm{gas}} \gtrsim 200 ~ \mathrm{K}$.

Combining high number densities and radiation fields, \hm\ formation on PAHs builds up a sufficiently high \hm\ column density at a relatively low total column density, such that \hm\ is already self-shielded against photodissociation.
At these low total column densities, dust temperatures are still too high for the formation of \hm\ on dust grains to have any effect ($T_{\mathrm{dust}} > 40$K).
The column density of the H/\hm\ transition is in that case entirely due to \hm\ formation on PAHs.

\section{Discussion}
\label{sec:discussion}

\subsection{Impact of uncertainties}
\label{ssec:impunc}
We find that PAHs, with coronene as a prototypical PAH, affect the structure of a PDR.
The formation of \hm\ on coronene is primarily important on the outskirts of the PDR.
In these regions the temperature is sufficiently high that the exact height of the barriers for addition and abstraction does not matter.
Deeper inside the cloud the barrier height does matter, but the transition to molecular hydrogen is then already complete, so \hm\ formation on PAHs is not significant in these regions.

Furthermore, most of the \hm\ formation on PAHs takes place via photodesorption.
These processes occur at low column densities, where the radiation field suffers almost no extinction.
At these column densities, photodesorption processes dominate and have a strong influence on the depth of the H/\hm\ transition.
To understand the dependence of the H/\hm\ transition on these photorates, we vary the binding energies of either H atoms or \hm\ molecules with 0.2 eV \textit{\textup{ceteris paribus}} and recalculate the photorates.
With these photorates the simulations are re-evaluated and show only a weak influence on the position of the H/\hm\ front.

Increasing the \hm\ or decreasing the H photodissociation energy ($E_0$ in Table \ref{tab:rates}) with 0.2 eV results in \hm\ formation rates reduced by a factor 3.
Conversely, the \hm\ formation rates locally increase by a factor 3 with a similar decrease in the photodissociation energy for \hm\ loss, or an increase in the photodissociation energy for H loss.
However, this does not significantly change the position of the H/\hm\ front.
Furthermore, we see in our simulations that in the clouds where PAHs have a strong influence on the cloud structure, their \hm\ formation rate is typically one order of magnitude higher than the dust grain \hm\ formation rate.

Another parameter that is subject of discussion is the cross sections used for the hydrogenation and abstraction reactions.
The reaction rate scales linearly with the cross section, and the values used in this paper are an average value derived from irradiating coronene films with H atoms \citep{mennella2012}.
\citet{mennella2012} argued that because coronene molecules only
weakly interact, and since the H atoms interact directly with the coronene molecules, without diffusion or absorption, their values are representative for the gas phase molecules as well.
In our model abstraction plays a minor role in \hm\ formation, but this changes for larger PAHs (Sect. \ref{ssec:imlar}).
In the cases where abstraction is the dominant mechanism for \hm\ formation, the formation rate linearly depends on both the abstraction cross section and the number of H atoms adsorbed onto PAHs.
The latter in turn depends on the cross section for the addition of an H atom.
We therefore expect that the rate of \hm\ formation through abstraction linearly depends on both the cross section for addition and abstraction.
To make more reliable \hm\ formation models for larger PAHs than this proof-of-concept approach, both these cross sections have to be established for larger PAHs.

\subsection{Expectations for a population of large PAHs}
\label{ssec:imlar}

We have used coronene as the only active PAH, where in reality there are many more PAHs of different sizes present in the ISM.
For PDR environments, PAHs with more than 40 carbon atoms and a compact structure are favoured because of their photostability.
\citet{allain1996a} showed that both H loss and \hm\ loss significantly decrease with increasing number of carbon atoms.
This makes \hm\ photodesorption a less likely process for \hm\ formation, but the decreasing rate of H loss will leave more H atoms on the PAH molecule, making \hm\ formation through abstraction a more dominant process.
This also changes the distribution of hydrogenation states throughout the PDR and increases the abundance of higher hydrogenation states.
These changes in both the dominant \hm\ formation mechanism on PAHs and the hydrogenation state distribution are observed in a test run of our model where coronene is substituted by a generic 50C PAH.
Since the chemistry between PAHs and hydrogen atoms is less well
understood for larger PAHs, we did not include these molecules in our simulations.

Larger molecules like circumcircumcoronene ($\mathrm{C_{96}H_{24}}$) are found to be in the highest hydrogenation state throughout the entire cloud \citep{montillaud2013}.
However, the photo-stability of hydrogenated PAHs is quite unknown, and preliminary experimental results suggest that the addition of hydrogen on PAHs changes their photostability \citep{reitsma2014}. 
This implies that the dissociation rates of hydrogenated coronene molecules could be lower than previously assumed.
As a result, the formation of \hm\ through photodesorption will decrease as well.
The photodesorption \hm\ formation rates presented in this paper are therefore an upper limit to possible \hm\ formation rates through this mechanism.
Simultaneously, \hm\ formation through abstraction will become more dominant on larger PAHs, since the decreased photodesorption rate leaves more H atoms on the PAH molecule.
However, to understand to which order abstraction can replace photodesorption as an \hm\ forming mechanism, a more comprehensive understanding of PAH (photo)chemistry is necessary.

Furthermore, there are other processes for \hm\ formation on PAHs, such as dissociative recombination between a superhydrogenated PAH cation and an electron \citep{lepage2009}.
This process is found to be effective for superhydrogenated PAHs with up to 200 carbon atoms.
In the environments we study the rate for this process is at least an order of magnitude lower than the dominant \hm\ formation processes.
To conclude about \hm\ formation on PAHs, it is necessary to fully understand every contributing process, since different processes appear to be the dominating mechanism for different PAH sizes and hydrogenation states.

\section{Conclusions}
\label{sec:conclusion}

The model we presented clearly shows that PAHs play a vital part in the location of the H/\hm\ transition in PDRs, especially at high gas temperatures ($T$ > 200 K).
For high UV fields ($G_{0} > 10^{3}$), PAH (photo)chemistry should therefore be taken into account when modelling the structure of a PDR.

We presented the first chemical model that accounts for more than one superhydrogenation of a PAH molecule and that incorporates thermal barriers into the PAH chemistry taking place.
This model also considers photodesorption as a mechanism for \hm\ formation on PAHs.
We showed that the barriers introduce a strong temperature dependence for the \hm\ formation route on PAHs, and that this formation pathway is the high-temperature complement to the classical \hm\ formation mechanism on dust grains.
Further research into PAHs should be directed at a better understanding of dehydrogenated PAHs and the chemical and photoprocesses that are associated with them.
Ideally, this is also directed towards larger PAHs, which are more representative of the ISM.

We conclude that photodesorption of \hm\ from PAHs can be an important pathway for the formation of \hm\ in PDR environments.
The barriers present in PAH hydrogenation limit the \hm\ formation process on PAHs to high gas temperatures.
These barriers therefore ensure that \hm\ formation on PAHs is the high-temperature complement to dust grain \hm\ formation.

\begin{acknowledgements}
L. B. and S. C. are supported by the Netherlands Organization for Scientific Research (NWO).
The authors would like to thank Seyit Hocuk for providing gas temperature models of PDRs.
The authors would also like to thank the anonymous referee for his or her helpful suggestions.
\end{acknowledgements}

\bibliographystyle{aa}

\bibliography{../../../boschman}

\pagebreak
\appendix
\section{Chemical reactions}

\begin{longtable}{lcccc}
\caption{\label{tab:reactions}Chemical reactions with different hydrogenation and charge states of coronene, as shown in the first column. The second and third column show the barrier and number of reactive carbon atoms for addition of a hydrogen atom. The fourth and fifth column show these properties for the process of abstraction, which leads to \hm\ formation.}\\
\hline\hline
 & \multicolumn{2}{c}{Hydrogenation} & \multicolumn{2}{c}{Abstraction} \\
Reactant                      &         Barrier / meV (K)                                  &  Reactive sites &   Barrier / meV (K) &   Reactive sites \\
\hline
\endfirsthead
\caption{continued.}\\
\hline\hline
 & \multicolumn{2}{c}{Hydrogenation} & \multicolumn{2}{c}{Abstraction} \\
Reactant                      &         Barrier / meV (K)                                  &  Reactive sites &   Barrier / meV (K) &   Reactive sites \\
\hline
\endhead
\hline
\endfoot

\hline
$\mathrm{C_{24}H_{0}}$ & - & 12 & - & - \\ 
$\mathrm{C_{24}H_{1}}$ & - & 11 & 10 (116) & 1 \\ 
$\mathrm{C_{24}H_{2}}$ & - & 10 & 10 (116) & 2 \\ 
$\mathrm{C_{24}H_{3}}$ & - & 9 & 10 (116) & 3 \\ 
\hline
$\mathrm{C_{24}H_{4}}$ & - & 8 & 10 (116) & 4 \\ 
$\mathrm{C_{24}H_{5}}$ & - & 7 & 10 (116) & 5 \\ 
$\mathrm{C_{24}H_{6}}$ & - & 6 & 10 (116) & 6 \\ 
$\mathrm{C_{24}H_{7}}$ & - & 5 & 10 (116) & 7 \\ 
\hline
$\mathrm{C_{24}H_{8}}$ & - & 4 & 10 (116) & 8 \\ 
$\mathrm{C_{24}H_{9}}$ & - & 3 & 10 (116) & 9 \\ 
$\mathrm{C_{24}H_{10}}$ & - & 2 & 10 (116) & 10 \\ 
$\mathrm{C_{24}H_{11}}$ & - & 1 & 10 (116) & 11 \\ 
\hline
$\mathrm{C_{24}H_{12}}$ & 28 (324) & 12 & 10 (116) & 12 \\ 
$\mathrm{C_{24}H_{13}}$ & - & 1 & - & 1 \\ 
$\mathrm{C_{24}H_{14}}$ & 46 (533) & 2 & 10 (116) & 2 \\ 
$\mathrm{C_{24}H_{15}}$ & - & 1 & - & 1 \\ 
\hline
$\mathrm{C_{24}H_{16}}$ & 64 (742) & 1 & 10 (116) & 4 \\ 
$\mathrm{C_{24}H_{17}}$ & - & 2 & - & 1 \\ 
$\mathrm{C_{24}H_{18}}$ & 35 (406) & 1 & 10 (116) & 6 \\ 
$\mathrm{C_{24}H_{19}}$ & - & 1 & - & 1 \\ 
\hline
$\mathrm{C_{24}H_{20}}$ & 30 (348) & 2 & 10 (116) & 8 \\ 
$\mathrm{C_{24}H_{21}}$ & - & 1 & - & 1 \\ 
$\mathrm{C_{24}H_{22}}$ & 52 (603) & 1 & 10 (116) & 10 \\ 
$\mathrm{C_{24}H_{23}}$ & - & 2 & - & 1 \\ 
\hline
$\mathrm{C_{24}H_{24}}$ & 33 (382) & 1 & 10 (116) & 12 \\ 
$\mathrm{C_{24}H_{25}}$ & - & 1 & - & 1 \\ 
$\mathrm{C_{24}H_{26}}$ & 33 (382) & 2 & 10 (116) & 14 \\ 
$\mathrm{C_{24}H_{27}}$ & - & 1 & - & 1 \\ 
\hline
$\mathrm{C_{24}H_{28}}$ & 39 (452) & 1 & 10 (116) & 16 \\ 
$\mathrm{C_{24}H_{29}}$ & - & 1 & - & 1 \\ 
$\mathrm{C_{24}H_{30}}$ & 34 (394) & 2 & 10 (116) & 18 \\ 
$\mathrm{C_{24}H_{31}}$ & - & 1 & - & 1 \\ 
\hline
$\mathrm{C_{24}H_{32}}$ & 49 (568) & 2 & 10 (116) & 20 \\ 
$\mathrm{C_{24}H_{33}}$ & - & 2 & - & 1 \\ 
$\mathrm{C_{24}H_{34}}$ & 49 (568) & 1 & 10 (116) & 22 \\ 
$\mathrm{C_{24}H_{35}}$ & - & 1 & - & 1 \\ 
\hline
$\mathrm{C_{24}H_{0}^{+}}$ & - & 12 & - & - \\ 
$\mathrm{C_{24}H_{1}^{+}}$ & - & 11 & 10 (116) & 1 \\ 
$\mathrm{C_{24}H_{2}^{+}}$ & - & 10 & 10 (116) & 2 \\ 
$\mathrm{C_{24}H_{3}^{+}}$ & - & 9 & 10 (116) & 3 \\ 
\hline
$\mathrm{C_{24}H_{4}^{+}}$ & - & 8 & 10 (116) & 4 \\ 
$\mathrm{C_{24}H_{5}^{+}}$ & - & 7 & 10 (116) & 5 \\ 
$\mathrm{C_{24}H_{6}^{+}}$ & - & 6 & 10 (116) & 6 \\ 
$\mathrm{C_{24}H_{7}^{+}}$ & - & 5 & 10 (116) & 7 \\ 
\hline
$\mathrm{C_{24}H_{8}^{+}}$ & - & 4 & 10 (116) & 8 \\ 
$\mathrm{C_{24}H_{9}^{+}}$ & - & 3 & 10 (116) & 9 \\ 
$\mathrm{C_{24}H_{10}^{+}}$ & - & 2 & 10 (116) & 10 \\ 
$\mathrm{C_{24}H_{11}^{+}}$ & - & 1 & 10 (116) & 11 \\ 
\hline
$\mathrm{C_{24}H_{12}^{+}}$ & - & 12 & 10 (116) & 12 \\ 
$\mathrm{C_{24}H_{13}^{+}}$ & 28 (324) & 1 & - & 1 \\ 
$\mathrm{C_{24}H_{14}^{+}}$ & - & 2 & 10 (116) & 2 \\ 
$\mathrm{C_{24}H_{15}^{+}}$ & 46 (533) & 1 & - & 1 \\ 
\hline
$\mathrm{C_{24}H_{16}^{+}}$ & - & 1 & 10 (116) & 4 \\ 
$\mathrm{C_{24}H_{17}^{+}}$ & 64 (742) & 2 & - & 1 \\ 
$\mathrm{C_{24}H_{18}^{+}}$ & - & 1 & 10 (116) & 6 \\ 
$\mathrm{C_{24}H_{19}^{+}}$ & 35 (406) & 1 & - & 1 \\ 
\hline
$\mathrm{C_{24}H_{20}^{+}}$ & - & 2 & 10 (116) & 8 \\ 
$\mathrm{C_{24}H_{21}^{+}}$ & 30 (348) & 1 & - & 1 \\ 
$\mathrm{C_{24}H_{22}^{+}}$ & - & 1 & 10 (116) & 10 \\ 
$\mathrm{C_{24}H_{23}^{+}}$ & 52 (603) & 2 & - & 1 \\ 
\hline
$\mathrm{C_{24}H_{24}^{+}}$ & - & 1 & 10 (116) & 12 \\ 
$\mathrm{C_{24}H_{25}^{+}}$ & 33 (382) & 1 & - & 1 \\ 
$\mathrm{C_{24}H_{26}^{+}}$ & - & 2 & 10 (116) & 14 \\ 
$\mathrm{C_{24}H_{27}^{+}}$ & 33 (382) & 1 & - & 1 \\ 
\hline
$\mathrm{C_{24}H_{28}^{+}}$ & - & 1 & 10 (116) & 16 \\ 
$\mathrm{C_{24}H_{29}^{+}}$ & 39 (452) & 1 & - & 1 \\ 
$\mathrm{C_{24}H_{30}^{+}}$ & - & 2 & 10 (116) & 18 \\ 
$\mathrm{C_{24}H_{31}^{+}}$ & 34 (394) & 1 & - & 1 \\ 
\hline
$\mathrm{C_{24}H_{32}^{+}}$ & - & 2 & 10 (116) & 20 \\ 
$\mathrm{C_{24}H_{33}^{+}}$ & 49 (568) & 2 & - & 1 \\ 
$\mathrm{C_{24}H_{34}^{+}}$ & - & 1 & 10 (116) & 22 \\ 
$\mathrm{C_{24}H_{35}^{+}}$ & 49 (568) & 1 & - & 1 \\ 
\hline
$\mathrm{C_{24}H_{36}}$ & - & - & 10 (116) & 24 \\ 
$\mathrm{C_{24}H_{36}^{+}}$ & - & - & 10 (116) & 24 \\

\end{longtable}

\begin{table*}[h!]
\caption{Integrated photodissociation rates for the different hydrogenation and ionization states of the coronene molecule. $E_{0}$ is the energy necessary for the process and $\Delta S$ is the energy difference between the transition state and the initial molecule \citep{lepage2001}. Values for $E_{0}$ and $\Delta S$ are taken from \citet{lepage2001} and \citet{paris2014}.}
\label{tab:rates}
\centering
  
\begin{tabular}{l c c c}

  \hline\hline

  Reaction & $E_{0}$ (eV) & $\Delta S$ ($\mathrm{cal \ K^{-1}}$) & $k_{\mathrm{ph}}$ ($\mathrm{s^{-1}} \ G_{0}^{-1}$) \\
  
  \hline
  
  $\mathrm{Cor}$ + h$\nu \rightarrow \mathrm{CorH_{-1}}$ + H                         & 4.73 & 5.0 & $9.29 \cdot 10^{-10}$ \\
  $\mathrm{Cor}$ + h$\nu \rightarrow \mathrm{CorH_{-2}}$ + \hm\                         & 4.75 & 5.0 & $9.29 \cdot 10^{-10}$ \\

  $\mathrm{Cor^{+}}$ + h$\nu \rightarrow \mathrm{CorH_{-1}^{+}}$ + H             & 4.85 & 5.0 & $3.38 \cdot 10^{-10}$ \\
  $\mathrm{Cor^{+}}$ + h$\nu \rightarrow \mathrm{CorH_{-2}^{+}}$ + \hm\  & 4.82 & 5.0 & $3.38 \cdot 10^{-10}$ \\
  \hline
  
  $\mathrm{CorH_{-n}}$ + h$\nu \rightarrow \mathrm{CorH_{-(n+1)}}$ + H             & 3.56 & 5.0 & $3.49 \cdot 10^{-9}$ \\
  $\mathrm{CorH_{-n}}$ + h$\nu \rightarrow \mathrm{CorH_{-(n+2)}}$ + \hm\  & 3.56 & 5.0 & $3.49 \cdot 10^{-9}$ \\

  $\mathrm{CorH_{-n}^{+}}$ + h$\nu \rightarrow \mathrm{CorH_{-(n+1)}^{+}}$ + H     & 4.28 & 5.0 & $5.05 \cdot 10^{-9}$ \\
  $\mathrm{CorH_{-n}^{+}}$ + h$\nu \rightarrow \mathrm{CorH_{-(n+2)}^{+}}$ + \hm\ & 4.28 & 5.0 & $5.05 \cdot 10^{-9}$ \\
  \hline
  
  $\mathrm{CorH_{n}}$ + h$\nu \rightarrow \mathrm{CorH_{n-1}}$ + H             & 1.2 & 5.0 & $2.58 \cdot 10^{-8}$ \\
  $\mathrm{CorH_{n}}$ + h$\nu \rightarrow \mathrm{CorH_{n-2}}$ + \hm\          & 1.6 & 5.0 & $1.23 \cdot 10^{-9}$ \\
  
  $\mathrm{CorH_{n}^{+}}$ + h$\nu \rightarrow \mathrm{CorH_{n-1}^{+}}$ + H     & 2.9 & 5.0 & $2.52 \cdot 10^{-8}$ \\
  $\mathrm{CorH_{n}^{+}}$ + h$\nu \rightarrow \mathrm{CorH_{n-2}^{+}}$ + \hm\  & 3.2 & 5.0 & $4.91 \cdot 10^{-9}$ \\
  \hline
  \hline
\end{tabular}

\end{table*}

\end{document}